\documentclass[%
 aip,
 amsmath,amssymb,
 reprint,%
]{revtex4-2}

\usepackage{graphicx}
\usepackage{dcolumn}
\usepackage{bm}

\usepackage[utf8]{inputenc}
\usepackage[T1]{fontenc}
\usepackage{mathptmx}
\usepackage{etoolbox}

\makeatletter
\def\@email#1#2{%
 \endgroup
 \patchcmd{\titleblock@produce}
  {\frontmatter@RRAPformat}
  {\frontmatter@RRAPformat{\produce@RRAP{*#1\href{mailto:#2}{#2}}}\frontmatter@RRAPformat}
  {}{}
}%
\makeatother

\usepackage{color} 
\usepackage[normalem]{ulem}
\usepackage[mathlines, displaymath, right]{lineno}
\usepackage{hyperref}
\usepackage[japanese,english]{babel}

\begin{document}

\preprint{APS/123-QED}

\title{Wavy synchronization in locomotion of train millipedes}

\author{Momiji Yoshikawa}
\affiliation{College of Engineering Systems, University of Tsukuba, Ibaraki, Japan.}
\author{Ikkyu Aihara}%
\affiliation{
 Institute of Systems and Information Engineering, University of Tsukuba, Ibaraki, Japan.
}

\date{August 6, 2026}

\begin{abstract}
The sophisticated control of many legs by millipedes and centipedes establish well-organized temporal structures such as wavy synchronization in their gate patterns.
Quantification and mathematical modeling of such temporal patterns can contribute to the understandings of behavioral mechanisms in biological locomotion. 
In this paper, we investigated the temporal pattern and its mechanism in the locomotion of small millipedes, train millipedes (\textit{Parafontaria laminata armigera}), by combining empirical data with a system of coupled phase oscillators.
First, we performed behavioral experiments by using actual millipedes and characterized their locomotion patterns based on an order parameter capable of detecting wavy synchronization with a specific wavenumber $k$. 
Second, we proposed a phase oscillator model incorporating two phase shift parameters and estimated their suitable values by comparing numerical simulations of the model with the empirically derived wavenumber through the proposed order parameter. 
Consequently, we established a concise mathematical model reproducing not only the wavy synchronization of the train millipedes but also the other types of wavy synchronization.
From the biological points of views, our result suggests that the asymmetric interaction between neighboring legs is a key factor in the sophisticated motion control of the millipedes, enabling them to maintain a specific wavenumber during forward walking.
\end{abstract}

\maketitle

\begin{quotation}
Locomotion in many-legged animals, such as millipedes, is a marvel of biological coordination. 
To move forward seamlessly without stumbling, these animals must perfectly synchronize dozens of legs, establishing graceful, wave-like patterns. 
While mathematical models of synchronization have been extensively studied, quantifying the exact mechanisms of living millipedes remains unknown due to the scarcity of high-definition empirical data. 
In this study, we bridge the gap between biology and nonlinear dynamics by combining the video tracking of train millipedes with a coupled phase oscillator model.
We have revealed that the wavy synchronization of millipedes can be quantitatively explained by asymmetric interactions between neighboring legs, which is a fundamental mathematical rule that governs their forward motion. 
Our findings not only shed light on the general mechanisms orchestrating many-legged locomotion but also provide a concise mathematical framework for designing highly stable, bio-inspired robots.
\end{quotation}

\section{\label{sec:level1}Introduction}
Synchronization has been observed in various biological systems.
For instance, male fireflies aggregate and flash synchronously \cite{Buck1968, Copeland1994, Sarfati2020};
this behavior allows the males to effectively attract conspecific females.
Male frogs call alternately with neighbors through acoustic interaction \cite{Brush1989,Gerhardt2002, Wells2007, Aihara2014};
such anti-phase synchronization likely allows the males to advertise themselves towards conspecific females by avoiding call overlaps \cite{Gerhardt2002,greenfield2021}.
Thus, the observed synchronization patterns vary considerably depending on the animal species and plays an important role for animals to survive in the wild.

The mechanisms of synchronization have been well studied from a theoretical point of view.
Kuramoto proposed a mathematical model (the phase oscillator model) in which the periodicity and interaction of multiple elements are described by a time differential equation of the phases \cite{kuramoto1984}.
Theoretical and numerical studies have demonstrated that this model can explain various types of synchronization in various biological systems \cite{aihara2009, aihara2011}. 
In addition, recent advances of data-driven approaches based on a phase oscillator model emphasize the importance of parameter estimation to infer the behavioral mechanism of animals including human  \cite{ota2014, ota2020,arai2024interlimb,FURUKAWA202547}.

\begin{figure}[h]
\includegraphics[width=8truecm]{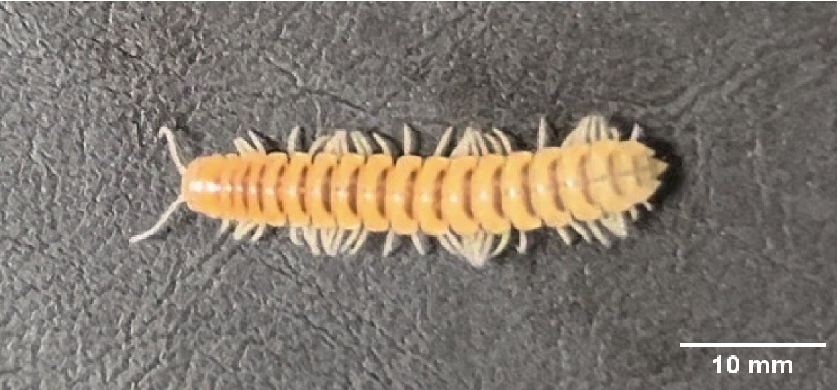}
\caption{\label{fig1} Photograph of a train millipede walking on a flat surface. The millipede controls the individual legs and establishes a stable walking pattern. Since the legs on both sides of the body are synchronized in almost in-phase, we focus on the dynamics of the legs on one side of the body. }
\end{figure}

Locomotion of many-legged animals such as millipedes and centipedes has been attracting a great deal of attention in physics and biology \cite{herreid2012locomotion}. 
Laboratory experiments have demonstrated the occurrence of wavy synchronization in multiple species of many-legged animals \cite{garcia2020fundamental,aoi2013instability,yasui2022adaptive,kano2017decentralized}.
However, quantitative studies based on high-definition video recordings have mainly been limited to several large-bodied species  such as \textit{Narceus americanus}, \textit{Spirostreptus giganteus}, and \textit{Scolopendra subspinipes}\cite{garcia2020fundamental,kano2017decentralized,yasui2022adaptive}.
Given that the previous studies have indicated the effect of body size on the temporal gait patterns \cite{garcia2020fundamental}, comprehensive analyses on multiple species (especially small-bodied millipedes and centipedes) are essential.
This study aims to examine the temporal structure and its mechanism in small millipedes, train millipedes (\textit{Parafontaria laminata armigera}) \cite{niijima1988,niijima2021eight}. 
While the body size of train millipedes is about 30mm \cite{niijima1988}, the preliminary observation has indicated the graceful wave-like locomotion patterns (Figure \ref{fig1}) \cite{arob2025}. 
Furthermore, in this study we use a phase oscillator model with data-driven approach to examine the mechanism of millipedes' locomotion.
It should be noted that several physical models are proposed for the locomotion of millipedes and centipedes \cite{kano2017decentralized,garcia2020fundamental,diaz2023active,yasui2022adaptive} but applying mechanically detailed models to smaller species is often challenging due to the unknown body structures. 


This paper is organized as follows.
First, we recorded the walking behavior of train millipedes and quantified the state of synchronization by extending the Kuramoto order parameter (Section \ref{sec:level2}).
Second, we estimated unknown parameters of a phase oscillator model with asymmetric phase shift parameters by combining simulation of the model with the empirically derived wavenumber through the proposed order parameter and showed that the proposed framework can quantitatively explain wavy synchronization of train millipedes as well as a general class of wavy synchronization with various wavenumber (Section \ref{sec: level3}).

\section{\label{sec:level2}Experiments}
\subsection{\label{subsec:level2a}Materials and methods}
We performed a laboratory experiment to quantify the walking behavior of train millipedes (Figure \ref{fig1}) as the time series data of the phases.
In October 2023, we captured two males and seven females of the millipedes in Mt.Hachibuse, Nagano Prefecture, Japan.
The millipedes were transported to our laboratory at University of Tsukuba.
The body length was 31.08 mm for the males and 33.07 mm for the females on average.
We recorded each millipede walking on a stand with a video camera of a smartphone (iPhone 14 MPUY3J/A, Apple Inc.) at the sampling rate of 60 fps.

By analyzing the video data, we tracked multiple body parts of each walking millipede (see Figure \ref{fig2}(a)).
For this analysis, we used DeepLabCut \cite{mathis2018deeplabcut}, a software for automatic tracking based on deep learning.
Specifically, we tracked the tips of the legs on one side of the body (30 legs for males and 31 legs for females) as well as the head and telson.
Then, we converted the coordinates of the tips of the legs to those relative to the body axis and subtracted the center of the orbit for each leg according to the procedure of the Appendix \ref{AppendixA}.
Consequently, we obtained the circular orbit around the origin for each leg (see Figure \ref{fig3}(a)) corresponding to the periodic rotation during locomotion.

Next, we carefully checked the result of the automatic tracking by comparing it with the video data.
We found that (1) the tracking errors rarely occurred (about 2\% of all the time series data) and 
(2) the $1^{\text{st}}$ - $3^{\text{rd}}$ legs were not used for locomotion in either males or females.
Regarding the first aspect, we corrected the misestimated coordinates by averaging the coordinates of the preceding and following data samples.
As for the second aspect, the $1^{\text{st}}$ - $3^{\text{rd}}$ legs were mostly suspended in the air and did not show periodic rotation during locomotion unlike the other legs.
Since this study focuses on the rhythm and locomotion, we excluded the $1^{\text{st}}$ - $3^{\text{rd}}$ legs from the subsequent analysis.
Finally, the angle of the leg was successfully estimated for 28 legs in females and 27 legs in males. 
We describe the angle of the $n$th leg as $\theta_n$ ranging from $-\pi$ to $\pi$.
This angle $\theta_n$ is defined to increase counterclockwise, starting from the front ($\theta_n = -\pi$), and progressing through touching down ($-\pi/2$), pushing back ($0$), lifting up ($\pi/2$) and returning to the front ($\pi$ or $-\pi$) (Figure \ref{fig3}(b)).

Let us introduce the criterion to quantify the degree of synchronization based on the Kuramoto order parameter.
The definition of the Kuramoto order parameter is as follows \cite{kuramoto1984}:
\begin{equation}
    r_{\text{in}} = \frac{1}{N} \left| \sum_{n=1}^N \exp{(i\theta_n)}\right|.
    \label{eq1}
\end{equation}
Here $N$ is the number of oscillators; $\theta_n \in [-\pi, \pi)$ represents the phase of the $n^{\text{th}}$ oscillator; $i$ is the imaginary unit.
This order parameter takes the maximum value of 1 when all the oscillators are synchronized with a phase difference of 0 \cite{kuramoto1984},
allowing us to detect the in-phase synchronization across all the oscillators.

To quantify the wavy pattern in walking millipedes, we extended the Kuramoto order parameter on the basis of a previous study \cite{Aihara2014} as follows:
\begin{equation}
    r_{\text{wavy}} = \frac{1}{N} \left| \sum_{n=1}^N \exp{\left\{i\left(\theta_n + \frac{2nk\pi}{N-1}\right)\right\}} \right|.
    \label{eq2}
\end{equation}
Here $k$ is the wavenumber; $\theta_n$ represents the angle of the $n^{\text{th}}$ leg.
Because of the gradual phase shift given by $2nk\pi/(N-1)$, this order parameter takes the maximum value of 1 when the wavy synchronization with the wavenumber $k$ occurs \cite{Aihara2014} (see Figure \ref{figA3} in the Appendix \ref{AppendixB} for details).
Note that (1) the absolute value of $k$ in this equation gives the number of wave cycles included in the body length and 
(2) the sign of $k$ describes the direction of the gradual phase shift which allows us to discriminate between forward wave and backward wave. 

Next, we estimated the wavenumber $k$ in Equation \eqref{eq2} from empirical data as follows:
\begin{enumerate}
    \item Based on video data, we determined a possible range of the wavenumber $k$. 
    \item On the assumption of a specific wavenumber $k$ within this range, we substituted the time series data of  $\theta_n$ into Equation \eqref{eq2} and calculated the average of the order parameter $r_{wavy}$ over time.
    \item We repeated the second step for the other possible values of $k$ within the range and determined the optimal wavenumber $k$ that had maximized the average of the order parameter $r_{wavy}$.
\end{enumerate}
Regarding a possible range of the wavenumber $k$ in the first step, we carefully checked video data and confirmed that approximately three waves were included in the gait pattern across all the legs (Figure \ref{fig2}(a)). 
Combined with the direction of the gradual phase shift in the wavy pattern of the millipedes, we determined the range to be $-3.5 \leq k \leq -2.5$. 
The above procedure was carried out for all the millipedes (two males and seven females in total). 

\subsection{\label{subsec:level2b}Results}
\begin{figure}[h]
\includegraphics[width=8truecm]{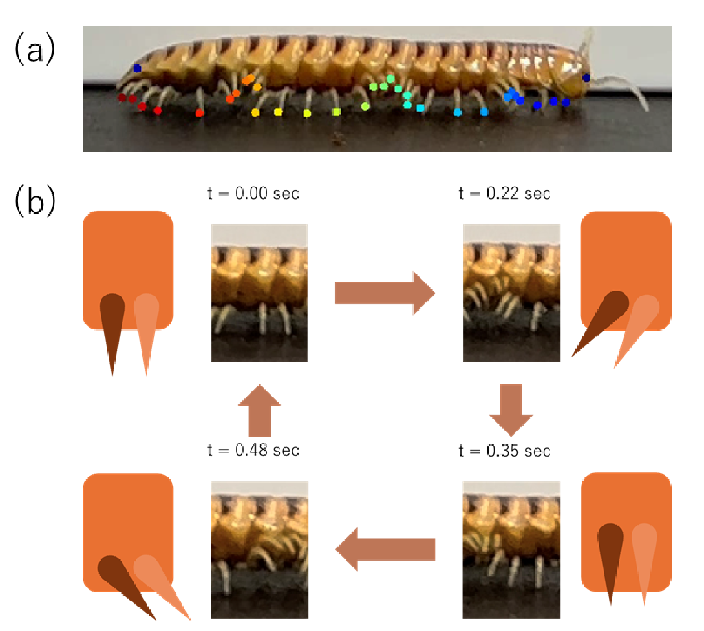}
\caption{\label{fig2} Automatic tracking of walking millipedes. 
(a) The tips of 31 legs with colored markers obtained from the analysis using DeepLabCut. 
(b) The rotation of specific legs in a walking millipede. When a millipede walked from left to right, these legs rotated clockwise.}
\end{figure}

The video recordings have revealed a wavy pattern in walking millipedes.
Figure \ref{fig2}(a) shows a snapshot of a walking millipede. 
Colored markers represent the tips of the legs that have been obtained from the automatic tracking via DeepLabCut, forming a wavy pattern as a whole. 
Figure \ref{fig2}(b) shows successive snapshots of neighboring legs. 
When the millipede walk from the left to the right, the legs involved in locomotion rotate clockwise.

\begin{figure*}
\includegraphics[width=16truecm]{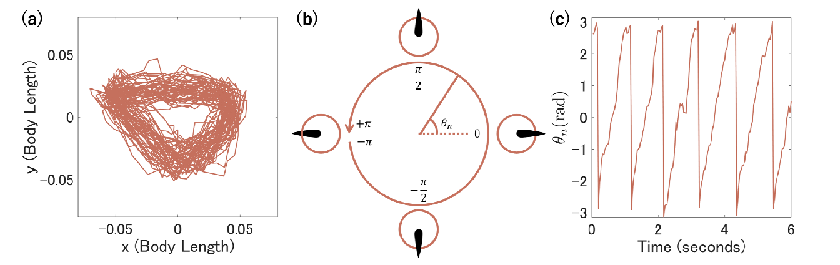}
\caption{\label{fig3} Empirical data on the phase dynamics of a representative leg during locomotion.
(a) Circular orbit of the leg relative to the body axis, which was obtained from automatic tracking.
(b) The definition of angle $\theta_n$. The black objects represent the positions of each leg at given angles.
(c) The time series data of the phase. The phase increases over time and is periodically reset to $-\pi$ when it hits $\pi$, corresponding to the periodic rotation of the leg. These graphs were obtained from the data of the $20^{\text{th}}$ leg of Female 4 (see Table \ref{table1}).
}
\end{figure*}

We have examined the dynamics of the rotating legs based on the phase $\theta_n$. 
Figures \ref{fig3}(a) and (c) show the circular orbit of a representative leg (the $20^{\text{th}}$ leg) and the time series data of its phase, respectively. 
$\theta_n$ increases over time and is periodically reset to $-\pi$, indicating strong periodicity in the locomotion. 
In our experiment, the walking direction was not fixed (in other words, some millipedes walked from the left to the right whereas the other millipedes walked from the right to the left). 
Our methodology allows us to define the phase $\theta_n$ that increases over time in both cases, by converting the coordinates of the legs relative to the body axis (see the Appendix \ref{AppendixA} for details). 
Figure \ref{fig4}(a) shows the phase dynamics of multiple legs (from the $15^{\text{th}}$ to the $20^{\text{th}}$ leg) that are displayed in different colors using a gradient. 
An almost constant phase shift has been observed between adjacent pairs of legs. 

Following the procedure in Section \ref{subsec:level2a}, we estimated the wavenumber $k$ from the empirical data. 
Figure \ref{fig4}(b) shows representative time series data of the order parameter $r_{wavy}$ for the estimated wavenumber $k=-2.95$ 
(corresponding to the result of Female 4 in Table \ref{table1}), 
demonstrating that $r_{wavy}$ maintained a high value close to 1. 
Figure \ref{fig4}(c) shows a snapshot of all phases at the median value of the estimated order parameter. 
The red lines indicate the angle of each leg and the circles schematically represent their circular orbits. 
This schematic figure is consistent with the wavy pattern of our empirical data (Figure \ref{fig2}(a)). 

Finally, we estimated the wavenumber $k$ for all collected millipedes (see Table \ref{table1}). 
The wavenumber $k$ was $-2.93\pm0.14$ (mean $\pm$ standard deviation) while the corresponding order parameter $r_{wavy}$ maintained a high value of $0.85\pm0.02$. 

\begin{figure}
\includegraphics[width=8truecm]{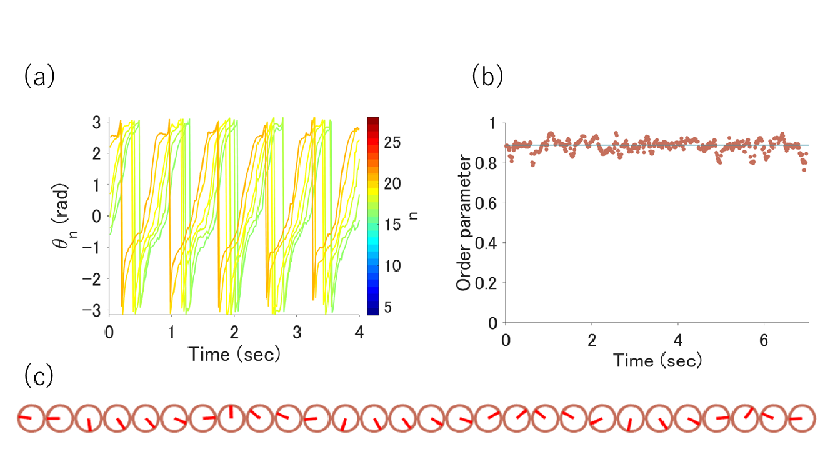}
\caption{\label{fig4} Wavy synchronization quantified from empirical data. 
(a) Phase dynamics of multiple legs (from the $15^{\text{th}}$ to the $20^{\text{th}}$ legs), visualized by a color gradient. 
(b) The time series data of the order parameter $r_{wavy}$ for the estimated wavenumber $k=-2.95$. 
The blue line represents the mean value. 
(c) The snapshot of the phases $\theta_n$ at the median value of the estimated order parameter, demonstrating the wavy pattern consistent with our observation. 
These graphs were obtained from the data of Female 4 (see Table \ref{table1}). }
\end{figure}

\begin{table}
    \begin{center}
        \caption{\label{table1}Quantification of wavy synchronization using the wavenumber $k$ in the extended order parameter $r_{wavy}$.}
        \begin{tabular}{r|l|l}
 & Order parameter (mean) & wavenumber \\ \hline
Male 1 & 0.84 & -3.03 \\
Male 2 & 0.86 & -3.04 \\
Female 1 & 0.86 & -3.08 \\
Female 2 & 0.83 & -3.01 \\
Female 3 & 0.87 & -3.05 \\
Female 4 & 0.89 & -2.95 \\
Female 5 & 0.86 & -2.78 \\
Female 6 & 0.86 & -2.78 \\
Female 7 & 0.82 & -2.68 \\ \hline
        \end{tabular}
    \end{center}
\end{table}

\section{\label{sec: level3}Mathematical Modeling}
\subsection{\label{subsec: level3a}Framework of the proposed model}
By extending the Kuramoto model, we propose a concise framework for the wavy synchronization of millipedes in the context of coupled phase oscillators. 
The definition of the Kuramoto model is as follows \cite{kuramoto1984}:
\begin{equation}
    \frac{d\theta_n}{dt} = \omega_n - \frac{K}{N}\sum_{m=1}^{N} \sin{(\theta_n - \theta_m)}.
    \label{eq3}
\end{equation}
Here $\theta_n \in [-\pi, \pi)$ is the phase of the $n^{\text{th}}$ oscillator; $\omega_n$ is the natural angular velocity of the $n^{\text{th}}$ oscillator; 
$N$ is the number of oscillators; $K$ is the coupling strength of the interaction between oscillators. 
Theoretical studies have shown that this model can qualitatively explain in-phase synchronization across all oscillators when the variance of $\omega_n$ is sufficiently smaller than the coupling strength $K$ \cite{kuramoto1984}.

As for the phase difference of adjacent oscillators, the wavy synchronization is similar to in-phase synchronization. 
Namely, the wavy synchronization can be described as the state in which the phases of adjacent oscillators are constantly shifted but their shift is small. 
Therefore, we propose a mathematical model of the walking millipedes by extending the Kuramoto model as follows: 
\begin{align}
    \frac{d\theta_1}{dt} &= \omega_1 - K\sin{(\theta_1 - \theta_2 + \alpha)}, \label{eq4}\\
    \frac{d\theta_n}{dt} &= \omega_n - \frac{K}{2} \left\{ \sin{(\theta_n - \theta_{n+1} + \alpha)} + \sin{(\theta_n - \theta_{n-1} + \beta)} \right\}, \nonumber \\
    & \hspace{10em} \text{(for $n = 2, \cdots, N-1$)} \label{eq5} \\ 
    \frac{d\theta_N}{dt} &= \omega_N - K\sin{(\theta_N - \theta_{N-1} + \beta)}. \label{eq6}
\end{align}
Here $\theta_n$ is the angle of the $n^{\text{th}}$ leg; $\omega_n$ is the natural angular velocity of the $n^{\text{th}}$ leg; $N$ is the number of legs involved in locomotion. 
In this model, we assume a sinusoidal interaction term with a coupling strength $K$ as in the Kuramoto model but introduce two phase shift parameters $\alpha$ and $\beta$ \cite{kori2001slow, aihara2009,sakaguchi1986soluble}. 
This formulation with a sinusoidal function corresponds to the first-order approximation of a Fourier expansion of a 2$\pi$-periodic function \cite{kuramoto1984} and also allows us to examine asymmetry in the interactions among adjacent legs.
The point is that millipedes usually move forward. 
To describe the asymmetry in the locomotion, we use  the two parameters $\alpha$ and $\beta$ and distinguish the effect of the behind leg (the ${n+1}^{\text{th}}$ leg) from that of the frontal leg (the ${n-1}^{\text{th}}$ leg).  
Finally, we normalize the strength of interaction by the number of legs with which the $n^{\text{th}}$ leg interacts. 

To estimate the suitable values of the phase shift parameters $\alpha$ and $\beta$, we fixed the remaining model parameters ($\omega_n$ and $K$), the wavenumber $k$ of the order parameter and initial conditions on the basis of (1) empirical data and (2) the dynamical features unique to a phase oscillator model. 
First, we assume that all legs share a time-independent and uniform natural angular velocity, $\omega_n = \omega$; 
this assumption has been widely adopted to mathematical modeling of many-legged animals \cite{yasui2017decentralized} because of the quite similar mechanical structure of the legs in the same species.
Then, we set the intrinsic angular velocity as $\omega = 2\pi/0.76$ [sec] based on empirical data of the rotating legs. 
Given the uniform natural angular velocity, $\omega_n = \omega$, the coupling strength $K$ merely scales the convergence time to the synchronized state without affecting the phase differences in the equilibrium state.
Therefore, we set $K=1.0$ without loss of generality.
Then, we set the wavenumber $k=-2.93$ which was obtained from our empirical results (see Section \ref{subsec:level2b} for details).
At last, the initial condition of the simulations was fixed at $\theta_n = -\pi/2$ [rad] based on our observation that the millipedes initially placed all their legs on the ground before starting to walk.

\subsection{\label{subsec: level3b}Results}
\begin{figure}
\includegraphics[width=8truecm]{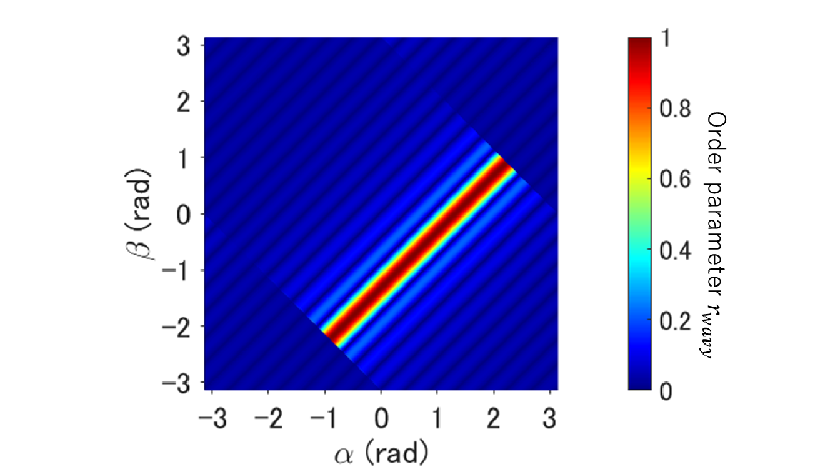}
\caption{\label{fig5} Numerical simulation to determine the phase shift parameters $\alpha$ and $\beta$ that have maximized the order parameter $r_{wavy}$ on the assumption of the wavenumber $k$ estimated from empirical data. Colors represent the value of $r_{wavy}$. 
The order parameter is maximized in the discontinuous linear region indicated in red. 
}
\end{figure}

We estimated the values of $\alpha$ and $\beta$ maximizing the order parameter $r_{wavy}$ 
on the assumption of the wavenumber estimated from empirical data. 
Specifically, we substituted the fixed parameters (i.e., $\omega_n$ and $K$) into Equations \eqref{eq4} - \eqref{eq6}, varied the remaining parameters $\alpha$ and $\beta$ in the range between $-\pi$ and $\pi$ at the interval of $0.001\pi$, 
and calculated the order parameter $r_{wavy}$ by numerically simulating the time evolution of Equations \eqref{eq4} - \eqref{eq6} on the assumption of the estimated wavenumber $k$. 
Figure \ref{fig5} shows how the order parameter $r_{wavy}$ depends on the values of $\alpha$ and $\beta$. 
It is demonstrated that $r_{wavy}$ is maximized in a discontinuous linear region indicated in red. 
Note that this result is consistent with a linear stability analysis of our model (see the Appendix \ref{AppendixB} for details). 

\begin{figure}
\includegraphics[width=8truecm]{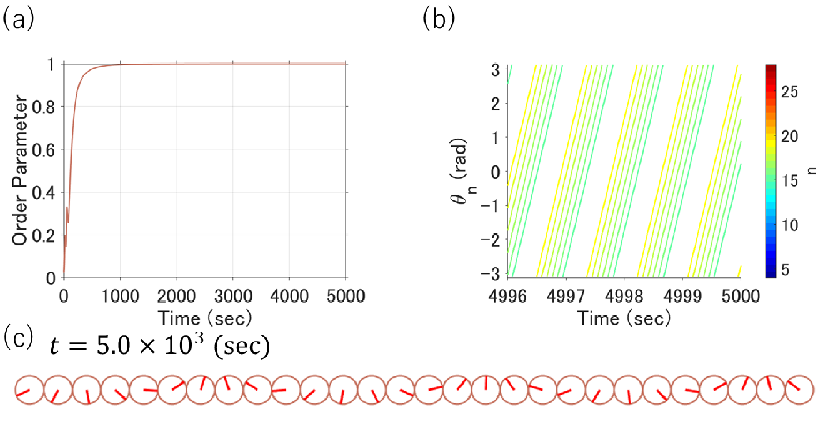}
\caption{\label{fig6} Numerical simulation of the proposed model with one set of the estimated parameter values ($\alpha = 0.679\pi$ and $\beta = 0.260\pi$). 
(a) The time evolution of the order parameter $r_{wavy}$. 
(b) The phase dynamics of multiple legs (from the $15^{\text{th}}$ to the $20^{\text{th}}$ leg) that are displayed in different colors using a gradient.
(c) A snapshot of the angles of all the legs. The wavy synchronization is achieved when the order parameter has converged to its maximum value of 1.
}
\end{figure}

Based on the result of Figure \ref{fig5}, we examined the detailed structure of the wavy synchronization produced by our model. 
Figure \ref{fig6} shows the result of numerical simulation with one set of the estimated parameter values ($\alpha=0.679\pi$ and $\beta=0.260\pi$). 
Figures \ref{fig6} (a) and (b) show that the wavy synchronization with a constant phase shift is achieved by these parameters.
Furthermore, a snapshot of the phases also shows the occurrence of the wavy synchronization (Figure \ref{fig6}(c)).
These results are consistent with empirically observed wavy synchronization (Figure \ref{fig4}(c)). 

\begin{figure}
\includegraphics[width=8truecm]{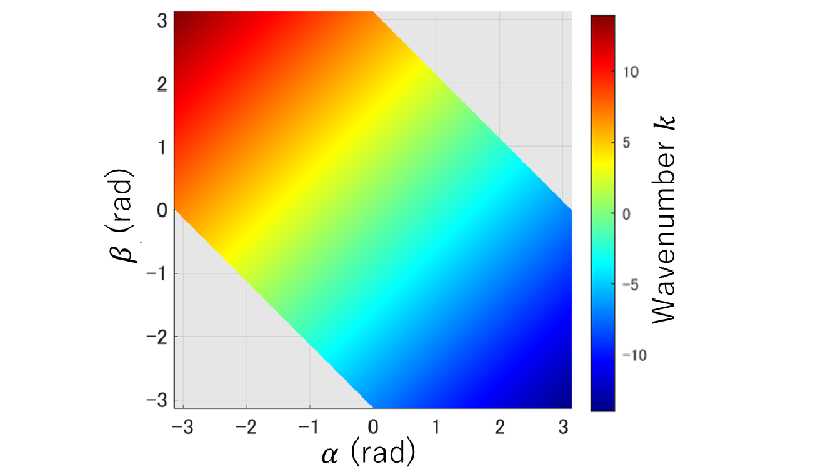}
\caption{\label{fig7}
Dependency of the wavenumber $k$ on the phase shift parameters $\alpha$ and $\beta$ within the parameter space
where the wavy synchronization is a stable equilibrium state.
The color gradient represents the value of the realized wavenumber $k$. 
The gray regions indicate the parameter spaces where the wavy synchronization is an unstable equilibrium state 
(see the Appendix for details). 
The wavy synchronization with various wavenumber emerges within the wide region of 
the asymmetric phase shift parameters $\alpha \neq \beta$, 
whereas the in-phase synchronization (corresponding to $k=0$) occurs along the diagonal line 
of the symmetric phase shift parameters $\alpha = \beta$. 
}
\end{figure}

To investigate the generality of the proposed framework, we have examined how the wavenumber $k$ depends on the phase shift parameters $\alpha$ and $\beta$ over the entire parameter space. 
Figure \ref{fig7} illustrates the dependency of the wavenumber $k$ on $\alpha$ and $\beta$ within the parameter region 
where the wavy synchronization is a stable equilibrium state.
Specifically, we have identified the parameter region of the stable solution based on the linear stability analysis (see the Appendix \ref{AppendixB} for details), and have numerically calculated the wavenumber $k$ of the stable wavy synchronization. 
This result demonstrates that the proposed model can explain various types of wavy synchronization 
with a wide range of wavenumber by tuning the phase shift parameters $\alpha$ and $\beta$. 
Notably, the in-phase synchronization, corresponding to the wavenumber $k=0$, is observed along the diagonal line where the parameters are symmetric (i.e., $\alpha = \beta$). 
This feature suggests that symmetric interactions between anterior and posterior legs lead to in-phase synchronization
and cannot reproduce the wavy synchronization of the millipedes. 
In contrast, the wavy synchronization with non-zero wavenumber, characteristic of the traveling waves in millipedes, 
emerges in the region where the phase shift parameters are asymmetric ($\alpha \neq \beta$). 
Thus, the proposed model provides a comprehensive explanation for diverse patterns of synchronization, 
ranging from in-phase synchronization to wavy synchronization with various wavenumber, 
by simply adjusting the asymmetry of the local interactions.

\section{\label{sec: level4}Summary and Discussion}
In this paper, we investigated the locomotion of train millipedes by combining empirical data and mathematical modeling.
First, we recorded the locomotion of train millipedes using a video camera and quantified the spatio-temporal pattern of the locomotion as wavy synchronization with a wavenumber $k$.
Second, we proposed a phase oscillator model with asymmetric phase shift parameters and estimated their values using the wavenumber $k$ obtained from the empirical data through the order parameter $r_{\text{wavy}}$.
Numerical simulations have demonstrated that (1) the proposed model can quantitatively explain the wavy synchronization of train millipedes
and (2) the model can explain a general class of wavy synchronization with various wavenumber.

Here, we discuss the contributions of this study from the viewpoint of locomotion in many-legged animals.
Wavy synchronization has been reported in various species of millipedes and centipedes \cite{herreid2012locomotion}.
However, quantitative studies based on high-definition video recordings have been limited to a few large-bodied species.
In this study, we quantified the walking pattern of small millipedes, train millipedes (\textit{Parafontaria laminata armigera}), as wavy synchronization with a wavenumber of $k = -2.93$.
This value is similar to the wavenumber of \textit{Apheloria virginiensis} (approximately 3) \cite{garcia2020fundamental} but significantly different from that of \textit{Spirostreptus giganteus} (approximately 6) \cite{kano2017decentralized}.
In addition, we identified two phase shift parameters of the model using the wavenumber estimated from the empirical data.
To our knowledge, this is a novel approach combining a phase oscillator model and empirical data via the extended order parameter defined in Equation (\ref{eq2}), contributing to the quantification of behavioral mechanisms in many-legged animals.
The numerical simulations and linear stability analysis of our model have provided a general result: 
wavy synchronization with various wavenumber including the value of train millipedes can be explained by our model under the assumption of asymmetric phase shift parameters (Figure \ref{fig7}; see also Appendix \ref{AppendixB}).
Combined with the fact that the locomotion of millipedes inherently exhibits asymmetry (i.e., they always move forward), the asymmetric interaction between neighboring legs is a key factor in the sophisticated motion control of various millipede species, enabling them to realize an stable wavenumber during forward walking.


Here we discuss the limitations and possible extensions of this study. 
First, we have used the order parameter $r_{wavy}$ for parameter estimation but other methods are possible 
\cite{aihara2009, aihara2011,miyazaki2006determination,miyazaki2006method,cestnik2017reconstructing}. 
For example, Ota and Aoyagi proposed the method to estimate the unknown parameters of a phase oscillator model from long time series data of the phases using a Bayesian approach \cite{ota2014}; their method was applied to several kinds of biological data such as frog choruses \cite{ota2020} and human locomotion \cite{arai2024interlimb, FURUKAWA202547}. 
It remains as a future issue to compare our result with the parameter estimation of other methods. 
Second, more precise modeling would be useful to describe the detailed mechanism in the locomotion of millipedes. 
For instance, millipedes have two pairs of legs per body segment in general. 
The interaction of legs within a body segment and that between neighboring segments would be different, indicating that the extension of the proposed model (e.g., the increase of model parameters) is worth examining. 
Third, previous studies on the other species of centipedes and millipedes have demonstrated that the inclination and complexity of terrain affect the locomotion pattern \cite{garcia2020fundamental,diaz2023active}. 
Since we have only used a flat terrain without a slope, additional experiments using various terrain conditions are worth testing for train millipedes. 

Future directions of this study include the application of the walking mechanism of the small millipedes (train millipedes) to the development of bio-inspired robots. 
The sophisticated structures and behaviors of animals have inspired the development of various kinds of technologies, including the development of  a snake-like \cite{wright2012design}, a millipede-like \cite{garcia2020fundamental}, a centipede-like \cite{osuka2019centipede} and a worm-like robot \cite{kim2013soft}. 
It is expected that such a bio-inspired robot is applied to various fields, such as inspection tasks \cite{bando2016sound}.  
In addition to the wavy synchronization, we have observed that train millipedes can maintain their body axis as a line with minimal lateral deviation (Figure \ref{fig2}(a)) although the body of the millipedes is very soft and flexible; 
this feature could facilitate the development of robots that can stably transport various supplies, for example. 
The proposed model enables us to simply imitate the wavy synchronization with various wavenumber including the case of train millipedes. 

\begin{acknowledgments}
We are grateful to T. Aoyagi, K. Kawaguchi, H. Furukawa and R. Yamaji for their valuable comments on this study. 
This study was partially supported by SECOM Science and Technology Foundation.
\end{acknowledgments}

\appendix
\setcounter{figure}{0}
\renewcommand{\thefigure}{A\arabic{figure}}
\setcounter{table}{0}
\renewcommand{\thetable}{A\arabic{table}}
\section{\label{AppendixA}Estimation of coordinates of legs relative to a body axis}
Here we explain how to estimate the coordinates of respective legs relative to a body axis. 
To stably calculate the angles of legs  ($\theta_n$), we need to reconstruct the circular orbits of respective legs from the video data. 
For this purpose, it is advantageous that the body axis of each train millipede remains straight during locomotion (see Figure \ref{fig2}(a)). 
Therefore, we use the coordinates of the head and telson to estimate the coordinates of legs relative to the body axis. 

In this analysis, we assume the following coordinates (pixel) at time t (sec) for each millipede: 
\begin{itemize}
    \item The coordinates of the tip of the $n^\text{th}$ leg: ($x_n (t),y_n (t)$)
    \item The coordinates of the head: ($x_{\text{head}} (t),y_{\text{head}} (t)$)
    \item The coordinates of the telson: ($x_{\text{telson}} (t),y_{\text{telson}} (t)$)
\end{itemize}
Note that these coordinates were already obtained from the automatic tracking shown in Figure \ref{fig2}(a) as colored markers. 

Next, we transform the coordinates of the legs and telson by subtracting the coordinates of the head as follows: 
\begin{align}
    x'_n(t) &= x_n(t) - x_{\text{head}}(t), \\
    y'_n(t) &= y_n(t) - y_{\text{head}}(t), \\
    x'_{\text{telson}}(t) &= x_{\text{telson}}(t) - x_{\text{head}}(t), \\
    y'_{\text{telson}}(t) &= y_{\text{telson}}(t) - y_{\text{head}}(t)
\end{align}
This procedure allows us to fix the position of the head at the origin. 
Then, we rotate the body axis to align it with the horizontal axis using a rotation matrix characterized by $\phi(t)$
and normalize the scale by the body length $L(t)$ as follows: 
\begin{equation}
\begin{bmatrix}
    x''_n(t) \\
    y''_n(t)
\end{bmatrix}
= \frac{1}{L(t)}
\begin{bmatrix}
    \cos{\phi(t)} & -\sin{\phi(t)} \\
    \sin{\phi(t)} & \cos{\phi(t)}
\end{bmatrix}
\begin{bmatrix}
    x'_n(t) \\
    y'_n(t)
\end{bmatrix}
\end{equation}
where 
\begin{align}
    L(t) &= (x'_{\text{telson}}(t)^2 + y'_{\text{telson}}(t)^2)^{1/2}, \\
    \phi(t) &= -\arctan\left(\frac{y'_{\text{telson}}(t)}{x'_{\text{telson}}(t)}\right). 
\end{align}
Through the above procedure, we obtained the coordinates of each leg as $(x''_n(t) , y''_n(t))$ relative to the body axis. 
Then, we calculated the centroid of $(x''_n(t) , y''_n(t))$ and subtracted it from $(x''_n(t) , y''_n(t))$. 
Consequently, we obtained the circular orbit of each leg around the origin (see Figure \ref{fig3}(a) for example).

\setcounter{figure}{0}
\renewcommand{\thefigure}{B\arabic{figure}}
\setcounter{table}{0}
\renewcommand{\thetable}{B\arabic{table}}
\section{\label{AppendixB}Linear stability analysis}
Here we explain a linear stability analysis \cite{strogatz2024nonlinear} of wavy synchronization realized in our model. 
First, we define the phase difference between adjacent legs as $\varphi_{n, n+1} =\theta_n - \theta_{n+1}$. 
From Equations \eqref{eq4} - \eqref{eq6} and the identical angular velocities $\omega_n = 2\pi/0.76$ ($n = 1, \cdots, N$) explained in the main manuscript, the dynamics of $\varphi_{n, n+1}$ are given as follows: 
\begin{widetext} 
\begin{align}
\frac{d\varphi_{1,2}}{dt} &=
    -\frac{K}{2}
    \Big[
        2\sin(\varphi_{1,2} + \alpha)
        - \sin(\varphi_{2,3} + \alpha)
        + \sin(\varphi_{1,2} - \beta)
    \Big],
    \label{eqA8} \\[1em]
\frac{d\varphi_{n,n+1}}{dt} &=
    - \frac{K}{2}
    \Big[
        \sin(\varphi_{n,n+1} + \alpha)
        - \sin(\varphi_{n-1,n} - \beta)
        - \sin(\varphi_{n+1,n+2} + \alpha)
        + \sin(\varphi_{n,n+1} - \beta)
    \Big],\nonumber \\
     & \hspace{25em} \text{(for $n = 2, \cdots, N-2$)} \label{eqA9} \\[1em]
\frac{d\varphi_{N-1,N}}{dt} &=
    - \frac{K}{2}
    \Big[
        \sin(\varphi_{N-1,N} + \alpha)
        - \sin(\varphi_{N-2,N-1} - \beta)
        + 2\sin(\varphi_{N-1,N} - \beta)
    \Big].
    \label{eqA10}
\end{align}
\end{widetext}
Using a constant phase shift $\Phi_0$, we define the wavy synchronization as $\varphi_{1, 2} = \cdots = \varphi_{N-1, N} = \Phi_0$. 
Substituting this condition into Equations \eqref{eqA8} -- \eqref{eqA10} yields 
\begin{align}
    \frac{d\varphi_{1, 2}}{dt}\Big|_{\varphi_{1, 2} = \cdots = \Phi_0} &= -\frac{K}{2}\{\sin(\Phi_0 + \alpha) + \sin(\Phi_0 - \beta)\}, \label{eqA11}\\
    \frac{d\varphi_{n, n+1}}{dt}\Big|_{\varphi_{1, 2} = \cdots = \Phi_0} &= 0, \nonumber \\ 
    & \hspace{5em} \text{(for $n = 2, \cdots, N-2$)} \label{eqA12} \\[1em]
    \frac{d\varphi_{N-1, N}}{dt}\Big|_{\varphi_{1, 2} = \cdots = \Phi_0} &= -\frac{K}{2}\{\sin(\Phi_0 + \alpha) + \sin(\Phi_0 - \beta)\}. \label{eqA13}
\end{align}
There are the following two conditions, each of which makes the right-hand sides of Equations \eqref{eqA11} -- \eqref{eqA13} to be 0:
\begin{align}
    \beta &= \alpha + 2\Phi_0 + 2M\pi, \label{eqA14}\\
    \beta &= \pi - \alpha + 2M\pi, \label{eqA15}
\end{align}
where $M$ is an integer.
This result means that the wavy synchronization is an equilibrium state when Equation \eqref{eqA14} or \eqref{eqA15} is satisfied. 

Next, we introduce the perturbation $\delta_n$ around the equilibrium state $\varphi_{n, n+1} = \varphi_{n, n+1}^*$ ($n = 1, \cdots, N-1$). 
The linear stability of the equilibrium state can be determined on the basis of the dynamics of the perturbation $\delta_n$. 
We describe the right-hand side of Equations \eqref{eqA8}--\eqref{eqA10} using the function $f_n(\varphi_{1, 2}, \dots , \varphi_{N-1, N})$  ($n = 1, \cdots, N-1$), 
substitute $\varphi_{n, n+1} = \varphi_{n, n+1}^* + \delta_n$ into $f_n(\varphi_{1, 2}, \cdots , \varphi_{N-1, N})$, 
and calculate the first-order approximation for the perturbation $\delta_n$.
Consequently, the dynamics of the perturbation $\delta_n$ are given as follows: 
\begin{widetext}
    \begin{equation}
    \renewcommand{\arraystretch}{1.5}
    \frac{d}{dt}
    \begin{pmatrix}
        \delta_1 \vphantom{\frac{\partial f}{\partial \phi}} \\
        \delta_2 \vphantom{\frac{\partial f}{\partial \phi}} \\
        \vdots \\
        \delta_n \vphantom{\frac{\partial f}{\partial \phi}} \\
        \vdots \\
        \delta_{N-2} \vphantom{\frac{\partial f}{\partial \phi}} \\
        \delta_{N-1} \vphantom{\frac{\partial f}{\partial \phi}}
    \end{pmatrix}
    =
    \left(
        \begin{array}{cccccc}
            \frac{\partial f_1}{\partial \varphi_{1,2}} & \frac{\partial f_1}{\partial \varphi_{2,3}} & & & & \text{\huge{0}} \\
            \frac{\partial f_2}{\partial \varphi_{1,2}} & \frac{\partial f_2}{\partial \varphi_{2,3}} & \frac{\partial f_2}{\partial \varphi_{3,4}} & & & \\
             & \ddots & \ddots & \ddots & & \\
             & & \frac{\partial f_n}{\partial \varphi_{n-1,n}} & \frac{\partial f_n}{\partial \varphi_{n,n+1}} & \frac{\partial f_n}{\partial \varphi_{n+1,n+2}} & \\
             & & \ddots & \ddots & \ddots & \\
             & & & \frac{\partial f_{N-2}}{\partial \varphi_{N-3,N-2}} & \frac{\partial f_{N-2}}{\partial \varphi_{N-2,N-1}} & \frac{\partial f_{N-2}}{\partial \varphi_{N-1,N}} \\
            \text{\huge{0}} & & & & \frac{\partial f_{N-1}}{\partial \varphi_{N-2,N-1}} & \frac{\partial f_{N-1}}{\partial \varphi_{N-1,N}}
        \end{array}
    \right)
    \begin{pmatrix}
        \delta_1 \vphantom{\frac{\partial f}{\partial \phi}} \\
        \delta_2 \vphantom{\frac{\partial f}{\partial \phi}} \\
        \vdots \\
        \delta_n \vphantom{\frac{\partial f}{\partial \phi}} \\
        \vdots \\
        \delta_{N-2} \vphantom{\frac{\partial f}{\partial \phi}} \\
        \delta_{N-1} \vphantom{\frac{\partial f}{\partial \phi}}
    \end{pmatrix}
    \label{eqA16}
\end{equation}
\end{widetext}
where $\frac{\partial f_n}{\partial \varphi}$ means the partial derivative of $f_n(\varphi_{1, 2}, \dots , \varphi_{N-1, N})$ at the equilibrium state of $\varphi_{n, n+1} = \varphi_{n, n+1}^*$ ($n = 1, \cdots, N-1$).
The matrix on the right-hand side of Equation \eqref{eqA16} is the Jacobian matrix that determines the linear stability of the equilibrium state depending on its eigenvalues \cite{strogatz2024nonlinear}. 
Substituting the definition of the wavy synchronization (i.e., $\varphi_{1, 2} = \cdots =  \varphi_{N-1, N}=\Phi_0$) and Equation \eqref{eqA14} into Equation \eqref{eqA16} yields the following Jacobian matrix: 
\begin{equation}
\frac{K}{2} \cos(\Phi_0  + \alpha)
\begin{pmatrix}
-3 & 1  &    & & 0 \\
1  & -2 & 1  & &  \\
   & \ddots & \ddots & \ddots &  \\
   &    &  1  & -2  & 1 \\
0  &     &   & 1 & -3
\end{pmatrix}
\label{eqA17}
\end{equation}
Similarly, substituting the definition of the wavy synchronization (i.e., $\varphi_{1, 2} = \cdots =  \varphi_{N-1, N}=\Phi_0$) and Equation \eqref{eqA15} into Equation \eqref{eqA16} yields the following Jacobian matrix: 
\begin{equation}
- \frac{K}{2}\cos(\Phi_0 + \alpha)
\begin{pmatrix}
1 & -1 &   &  & 0 \\
1 & 0 & -1 &  &   \\
  & \ddots & \ddots & \ddots &  \\
  &   & 1  & 0 & -1 \\
0 &   &    & 1 & -1
\end{pmatrix}
\label{eqA18}
\end{equation}

\begin{figure}
\includegraphics[width=8truecm]{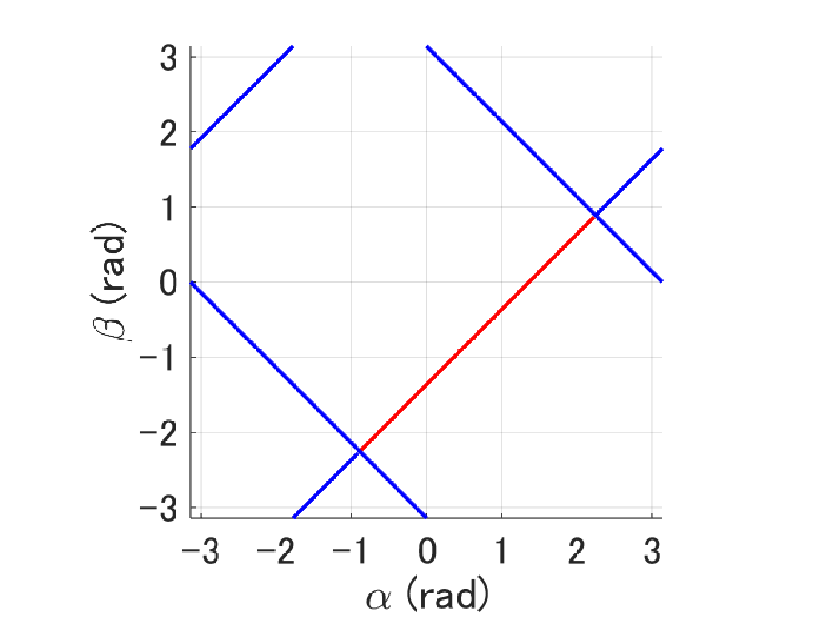}
\caption{\label{figA1} Linear stability analysis of the wavy synchronization on the assumption of the wavenumber $k=-2.93$ and the coupling strength $K=1.0$. The red line represents the parameter region in which the wavy synchronization is asymptotically stable; the blue lines represent the regions in which the wavy synchronization is asymptotically unstable or half-stable. }
\end{figure}

Next, we analyzed the linear stability of the wavy synchronization by using the eigenvalues of the Jacobian matrices 
(i.e., Equations \eqref{eqA17} and \eqref{eqA18}).  
Figure \ref{figA1} shows the parameter region in which the wavy synchronization 
with the constant phase shift $\Phi_0=2\pi k/(N-1)$ is an equilibrium state, 
on the assumption of the coupling strength $K=1.0$ and the wavenumber $k=-2.93$ that was estimated from the empirical data.
Note that the wavy synchronization is an equilibrium state along the lines of Equations \eqref{eqA14} and \eqref{eqA15}  while its stability is determined by the eigenvalues of the Jacobian matrices (Equations \eqref{eqA17} and \eqref{eqA18}).
The red line represents the region in which the wavy synchronization is asymptotically stable (in other words, the maximum of the real part of the eigenvalues is negative); 
the blue lines represent the regions in which the wavy synchronization is asymptotically unstable or half-stable (in other words, the maximum of the real part of the eigenvalues is larger than or equal to zero). 
The region of this red line is consistent with the strong red band of Figure \ref{fig5} in which the order parameter $r_{wavy}$ is maximized.

\begin{figure}
\includegraphics[width=8truecm]{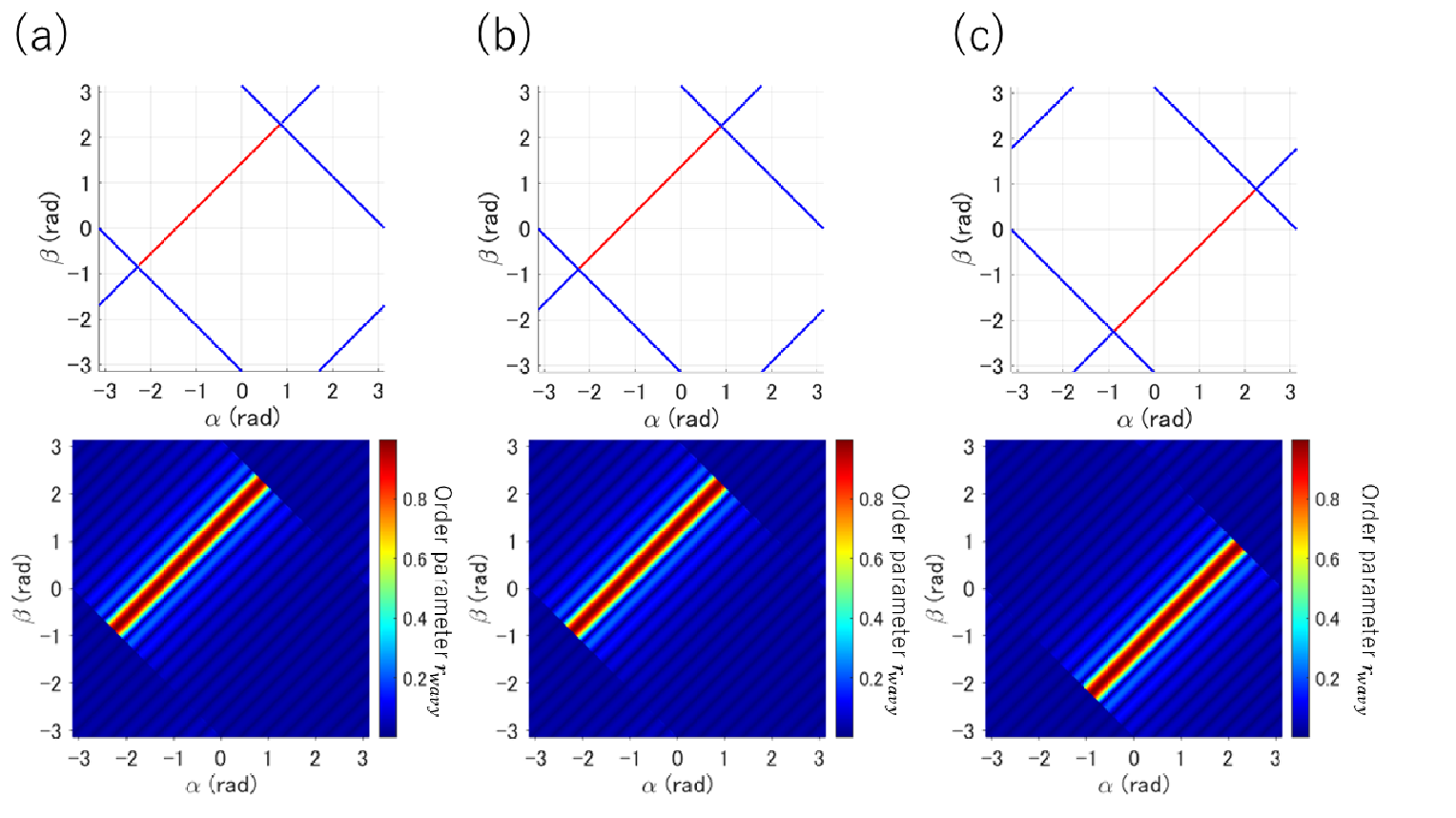}
\caption{\label{figA2} 
Wavy synchronization with various wavenumbers: (a) $k=3.10$, (b) $k= 2.93$ and (c) $k=-2.93$. 
The top panels represent the result of the linear stability analysis. 
Here, the red line represents the parameter region in which the wavy synchronization is asymptotically stable; 
the blue lines represent the regions in which the wavy synchronization is asymptotically unstable or half-stable. 
In contrast, the bottom panels represent the order parameter $r_{wavy}$ that we numerically calculated according to the method explained in the main manuscript
by varying the phase shift parameters $\alpha$ and $\beta$ at the interval of $0.001\pi$.
We have obtained the consistent results from these two kinds of analyses. }
\end{figure}

\begin{figure}
\includegraphics[width=8truecm]{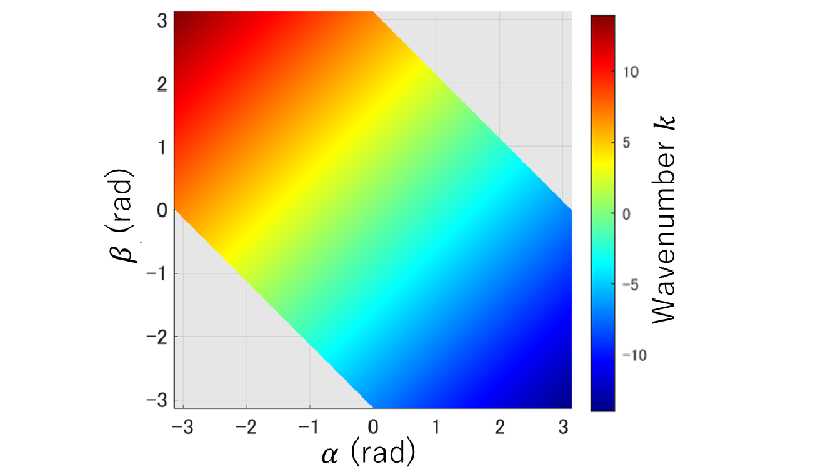}
\caption{\label{figB5} 
The dependency of the wavenumber $k$ on the phase shift parameters $\alpha$ and $\beta$. 
A color gradient describes the value of the wavenumber within the parameter region where the wavy synchronization exists as a stable equilibrium state. 
This figure has demonstrated that the wavy synchronization exists as a stable equilibrium state across the wide parameter region 
and its wavenumber gradually changes depending on $\alpha$ and $\beta$.
}
\end{figure}

\begin{figure}
\includegraphics[width=8truecm]{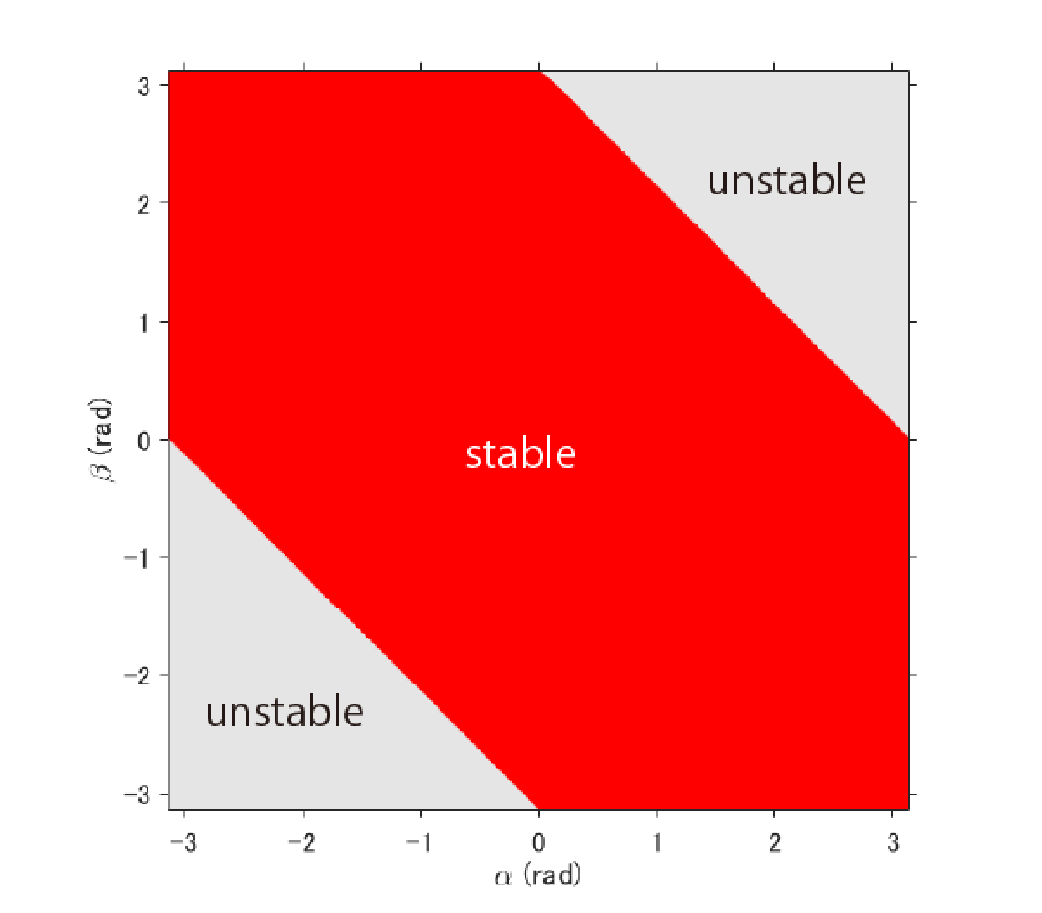}
\caption{\label{figB4} 
The visualization of the parameter regions in which the wavy synchronization exists as a stable equilibrium state or an unstable equilibrium state. 
We can confirm that there is the boundary where the linear stability of the wavy synchronization switches.
}
\end{figure}

By using the result of the linear stability analysis (Equations (\ref{eqA14}), (\ref{eqA15}), (\ref{eqA17}) and (\ref{eqA18})), 
we quantified how the wavenumber of the stable wavy synchronization depends on the phase shift parameters $\alpha$ and $\beta$. 
Top panels of Figure \ref{figA2} represent the results of the linear stability analysis on the assumption of different wavenumbers: (a) $k=3.10$, (b) $k= 2.93$ and (c) $k=-2.93$. 
Namely, we assumed a specific wavenumber $k$ among the three values, 
used Equations (\ref{eqA14}) and (\ref{eqA15}) to estimate the parameter region in which the wavy synchronization with the wavenumber $k$ is an equilibrium state, 
and calculated the linear stability based on the eigenvalues of the Jacobian matrices (Equations \eqref{eqA17} and \eqref{eqA18}).
It is demonstrated that the region of the stable solution indicated in red has changed depending on the assumed wavenumber. 
Moreover, we carried out more comprehensive analysis of how the wavenumber $k$ depends on the phase shift parameters across the entire parameter space.
In Figure \ref{figB5}, a color gradient describes the value of the wavenumber within the parameter region where the wavy synchronization exists as a stable equilibrium state. 
This figure has shown that the wavy synchronization exists as a stable equilibrium state across the wide parameter region 
and its wavenumber gradually changes depending on $\alpha$ and $\beta$.
Note that this result of the linear stability analysis is consistent with Figure \ref{fig7} in the main manuscript. 
We also show the parameter regions in which the wavy synchronization exists as a stable equilibrium state or  an unstable equilibrium state (Figure \ref{figB4}), 
allowing us to check the boundary where the linear stability of the wavy synchronization switches.
\begin{figure}
\includegraphics[width=8truecm]{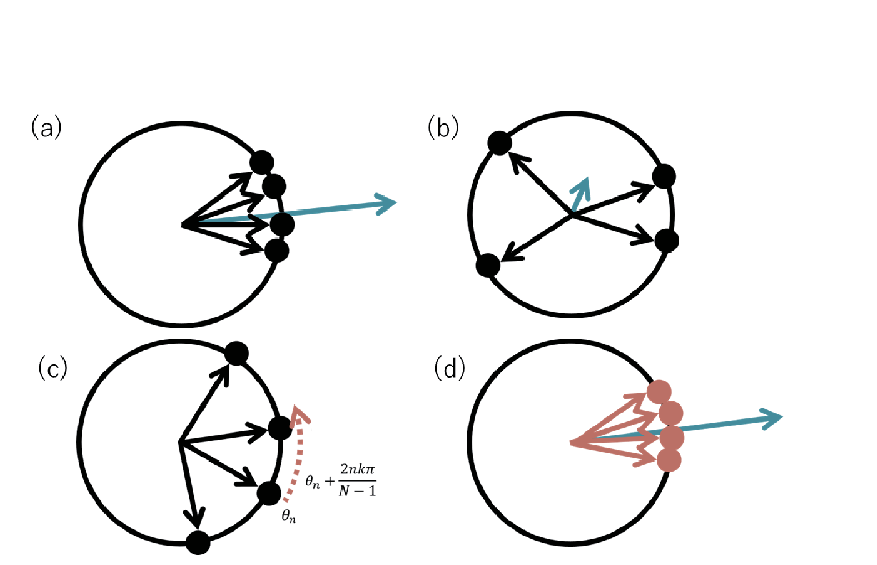}
\caption{\label{figA3} Schematic diagram of the order parameter used in this study. 
(a, b) The Kuramoto order parameter $r_{in}$ for the localized distribution and the scattered distribution of $\theta_n$. 
The phase $\theta_n$ is treated as a vector on a unit circle on the complex plane (i.e., $\exp(i\theta_n)$) . 
Then, the order parameter $r_{in}$ is given as the length of the mean vector (i.e., $|\sum_n(\exp(i\theta_n))|/N$). 
The point is that the resultant vector (i.e., $\sum_n(\exp(i\theta_n))$; a light blue arrow) becomes longer and $r_{in}$ takes a larger value when $\theta_n$ is localized (Figure \ref{figA3}(a)).  
In contrast, the resultant vector becomes shorter and $r_{in}$ takes a smaller value when $\theta_n$ is randomly distributed (Figure \ref{figA3}(b)). 
(c, d) The extended order parameter $r_{wavy}$. As shown in Equation \eqref{eq2}, $r_{wavy}$ includes the term $2nk\pi/(N-1)$ that describes the phase shift among oscillators. 
This term allows us to cancel the constant phase shift originating from the wavy synchronization (Figure \ref{figA3}(c)).
Consequently, the resultant vector becomes longer and $r_{wavy}$ takes a larger value (Figure \ref{figA3}(d)) when the wavy synchronization with the wavenumber $k$ occurs.
}
\end{figure}

\clearpage

\nocite{*}
\bibliography{aipsamp}

\end{document}